\documentclass[fleqn,usenatbib]{mnras}

\usepackage{newtxtext,newtxmath}

\usepackage[T1]{fontenc}

\DeclareRobustCommand{\VAN}[3]{#2}
\let\VANthebibliography\thebibliography
\def\thebibliography{\DeclareRobustCommand{\VAN}[3]{##3}\VANthebibliography}

\def\ks{km s$^{-1}$}
\def\d{$^\circ$}
\def\m{$^\prime$}
\def\s{$^{\prime\prime}$}
\def\hh{$^{\mathrm h}$}
\def\mm{$^{\mathrm m}$}
\def\ss{$^{\mathrm s}$}
\def\hii{H\textsc{ii}}
\def\msol{M$_\odot$}

\usepackage{graphicx}	
\usepackage{amsmath}	
\usepackage{amssymb}	

\title[Linking three generations of stars]{Three generations of stars: a possible case of triggered star formation}

\author[M. B. Areal et al.]{
M. B. Areal,$^{1}$\thanks{E-mail: mbareal@iafe.uba.ar (MBA)}
A. Buccino,$^{1,2}$
S. Paron,$^{1}$
C. Fari\~{n}a$^{3,4}$
and M. E. Ortega$^{1}$
\\
$^{1}$CONICET-Universidad de Buenos Aires. Instituto de Astronom\'ia y F\'isica del Espacio CC 67, Suc. 28, 1428 Buenos Aires, Argentina\\
$^{2}$Departamento de F\'isica. FCEyN-Universidad de Buenos Aires, Buenos Aires, Argentina.\\
$^{3}$Isaac Newton Group of Telescopes, E38700, La Palma, Spain.\\
$^{4}$Instituto de Astrof\'isica de Canarias (IAC) and Universidad de La Laguna, Dpto. Astrof\'isica, Spain.
}

\date{Accepted XXX. Received YYY; in original form ZZZ}

\pubyear{2020}

\begin{document}
\label{firstpage}
\pagerange{\pageref{firstpage}--\pageref{lastpage}}
\maketitle

\begin{abstract}

Evidence for triggered star formation linking three generations of stars is 
difficult to assemble, as it requires convincingly associating evolved massive stars 
with \hii~regions that, in turn, would need to present signs of active star formation. 
We present observational evidence for triggered star formation relating three generations 
of stars in the neighbourhood of the star LS~II~+26~8. We carried out new spectroscopic 
observations of LS~II~+26~8, revealing that it is a \textsc{B0 III}-type star.
We note that LS~II~+26~8 is located exactly at the geometric centre of a semi-shell-like 
\hii~region complex. The most conspicuous component of this complex is the \hii~region 
Sh2-90, which is probably triggering a new generation of stars. The distances to LS~II~+26~8 and to Sh2-90 
are in agreement (between 2.6 and 3 kpc). Analysis of the 
interstellar medium on a larger spatial scale shows that \hii~region complex lies on
the northwestern border of an extended H$_{2}$ shell. The radius of this molecular 
shell is about 13 pc, which is in agreement with what an O9V star (the probable initial 
spectral type of LS~II~+26~8 as inferred from evolutive tracks) can generate through its 
winds in the molecular environment. In conclusion, the spatial and temporal 
correspondences derived in our analysis enable us to propose a probable triggered star 
formation scenario initiated by the evolved massive star LS~II~+26~8 during its main sequence 
stage, followed by stars exciting the \hii~region complex formed in the molecular 
shell, and culminating in the birth of YSOs around Sh2-90.

\end{abstract}




\section{Introduction}

Triggered star formation linking two generations of stars has been studied, 
mainly at the interfaces between \hii~regions and molecular clouds.  It is usually 
observed that the older generation, i.e.\ the exciting stars of an \hii~region, 
triggers the formation of a new generation of stars at the \hii~regions's periphery 
(e.g.\ \citealt{duro17,deha15,zav10,zav07}). The exciting stars of the \hii~regions 
of course also have a formation history, but it is very difficult to find any evidence for it 
because of the complexity of the interstellar medium (ISM) and the time scales involved. 
Thus, evidence for triggered star formation linking three generations of stars is 
difficult to assemble, as it requires convincingly associating evolved massive stars with HII regions that,
in turn, would need to present signs of active star formation.
Indeed, observational evidence is scant and in general involves superstructures 
(e.g.\ supershells) in the ISM. For instance, \citet{oey05} argue that they identified the 
first example of three-generation, hierarchical triggered star formation in the W3/W4 Galactic 
star-forming complex. \citet{parker92}, \citet{wal92}, and 
\citet{barba03} have also suggested a case of possible sequential star formation among three 
generations of OB associations towards the N11B nebula in the Large Magellanic Cloud.

In the followings sections we present observational evidence pointing to a likely 
Galactic case of triggered star formation among three generations of stars.

\section{Presentation of the case}\label{sec.present}

\begin{figure}
    \centering
	\includegraphics[width=8cm]{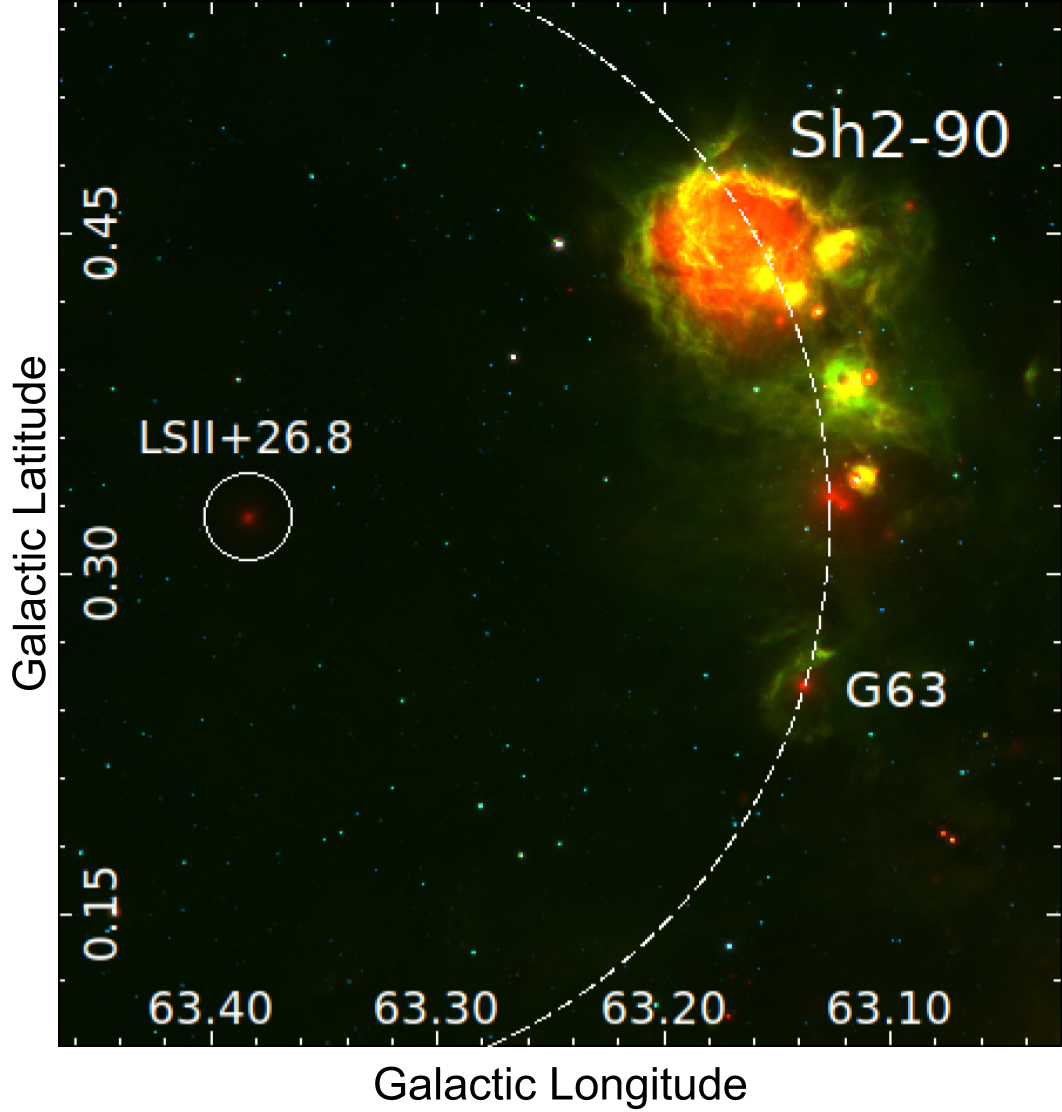}
    \caption{Three-colour image displaying the GLIMPSE/{\it Spitzer} emissions at 8 and 4.5~$\mu m$ 
(green and blue, respectively), and the MIPS/{\it Spitzer} at 24~$\mu m$ (red). It is noted 
that the star LS~II~+26~8 is located at the geometric centre of a semi-shell-like \hii~region
 complex, composed of Sh2-90 and HRDS G063.137+00.252 (indicated as G63).}
    \label{fig:Sh2-90}
\end{figure}

The star LS~II~+26~8 ($\alpha_{2000}$$=$19\hh50\mm12.7\ss, $\delta_{2000}$$=+$26\d58\m35\s) was catalogued, 
based on photometric criteria, as a possible OB-type star 
\citep{1960LS....C02....0S}. \citet{2018AJ....156...58B} obtained the parallax of LS~II~+26~8 and
estimated a distance of between 2.8 and 3.5 kpc. Inspection of the ISM surrounding the 
star reveals an interesting configuration. Figure\,\ref{fig:Sh2-90} is a three-colour 
image showing the 8 and 4.5~$\mu$m emission extracted from the GLIMPSE/{\it Spitzer} 
survey in green and blue, respectively, and 24~$\mu m$ obtained from the MIPS/{\it Spitzer} 
survey in red. It can be seen that LS~II~+26~8 is located exactly at the geometric 
centre of a semi-shell-like H\textsc{ii}~region complex (with a radius of about 15\m). 
Are the LS~II~+26~8 star and the \hii~region complex related?

The semi-shell-like H\textsc{ii}~region complex consists of a set of \hii~regions belonging to 
Sh2-90 \citep{1959ApJS....4..257S} and the \hii~region HRDS G063.137+00.252 \citep{2012ApJ...759...96B}, 
marked as G63 in Fig.\,\ref{fig:Sh2-90}. These are related to a molecular cloud that extends 
from 15.5 to 28.5~km s$^{-1}$~with a mean velocity of about 20.8  \ks~\citep{1983A&A...124....1L,2014A&A...566A.122S}. 
The latter authors, using the galactic rotation curve of \citet{brand93} to convert the 
mean velocity of the molecular cloud to distance and other indirect distance estimation methods, 
suggest that the distance to Sh2-90 lies in the range 2.1--2.5 kpc. This distance is 
somewhat close to the lower value of the distance to LS~II~+26~8 determined from parallax. 
In addition, if the galactic rotation curve of \citet{fich89} is used, the mean velocity of 
the molecular gas has an associated distance of about 3 kpc, in close agreement with the distance estimated for LS~II~+26~8.   
Moreover, using the revised kinematic distance estimate based on the method described in 
\citet{reid09} (Section 4)\footnote{bessel.vlbi-astrometry.org/revised\_kd\_2014} together with the 
Galactic parameters from the A5 model of \citet{reid14}, we obtain a distance of $2.5\pm1.0$ kpc 
for Sh2-90. In conclusion, we suggest that it is very likely that the star LS~II~+26~8 and the 
\hii~region complex are located closed enough to be physically related.

Sh2-90, which has an age $< 6.5 \times 10^{6}$ yr \citep{2014A&A...566A.122S}, is probably 
excited by a loose cluster of massive stars, of which the most massive member is an O8--O9V 
star. According to the authors, Sh2-90 is very probably triggering star formation. More than 100 
young stellar object (YSO) candidates with masses in the range 0.2--3 \msol~were found in its neighbourhood.

Given that this is a promising region to relate a possible evolved massive star with 
\hii~regions that are triggering star formation, and that we were probably registering a case of 
triggered star formation among three generations of stars, we performed spectroscopic optical 
observations to get an accurate determination of the spectral type of LS~II~+26~8 and investigate the 
surrounding ISM along the circumference shown in Fig.\,\ref{fig:Sh2-90}.

\section{Observations}

The visible spectrum of the star LS~II~+26~8 was obtained with the Intermediate Dispersion 
Spectrograph (IDS) mounted on the 2.5 m Isaac Newton Telescope at Roque de Los Muchachos 
Observatory on 2019 October 22. We employed the R900V grating, which allows us to obtain 
the spectrum in the 3800--5200~\AA~wavelength range with a mean dispersion of 0.69~\AA~per pixel.

Two consecutive images were obtained to filter out the cosmic rays in the combined spectrum. 
The images were  bias-corrected and the spectrum was optimally extracted and wavelength calibrated
using standard IRAF\footnote{Image Reduction and Analysis Facility (IRAF) is distributed by 
the National Optical Astronomy Observatories, which is operated by the Association of Universities 
for Research in Astronomy, Inc.,under contract to the National Science Foundation.} routines. 
The wavelength calibration was derived from the CuAr+CuNe arc-lamp spectra obtained at target 
position.The rms values of the wavelength calibration were less than 0.5~\AA. The resulting 
spectrum has a high S/N ($\sim$200) at 4500~\AA. The spectrum was normalized using standard IRAF tasks.

\section{Results}

\subsection{LS~II~+26~8 star: spectral classification}

\begin{figure*}
	\centering
    \includegraphics[width=17cm]{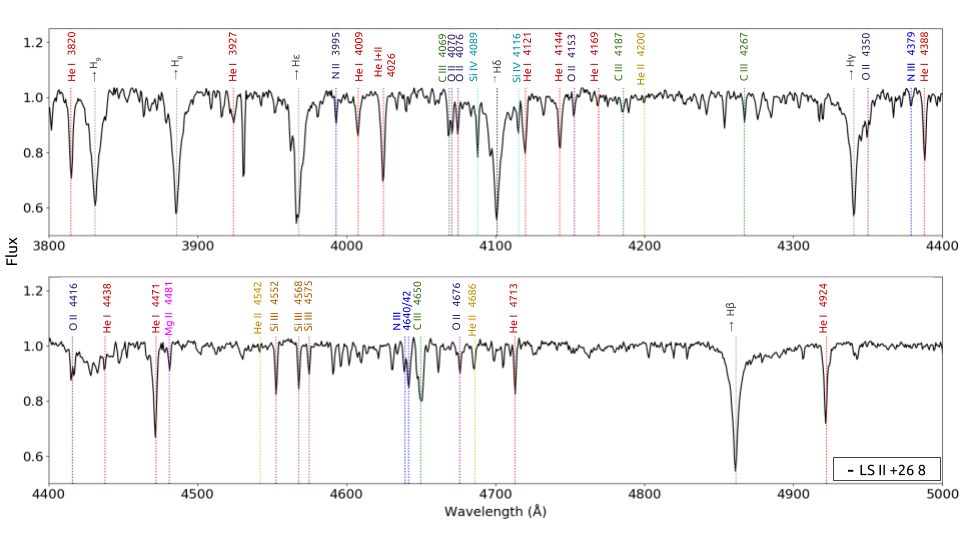}
    \caption{Spectrum of the LS~II~+26~8 obtained with the IDS at the Isaac Newton Telescope.}
    \label{fig:espectro}
\end{figure*}

Figure\,\ref{fig:espectro} presents the spectrum of LS~II~+26~8 obtained using the 
IDS, and Table\,\ref{tab:eqwidth} lists several of the observed lines that are important for 
the spectral classification from their equivalent widths (EWs). 

The LS~II~+26~8 spectrum presents prominent He~\textsc{i} absorption lines. He~\textsc{ii} 
lines can be identified, although very weak, except for He~\textsc{ii} 4686~\AA. According to 
the criteria of spectral classification for OB stars established by \cite{wal90}, He~\textsc{ii} 
4686~\AA~is last seen in the B0.5--B0.7 range of spectral types, whereas He~\textsc{ii} 4542~\AA~$\sim$ Si~\textsc{iii} 
4552~\AA~defines the O9.7 type. This, together with the presence of Mg~\textsc{ii} 4481~\AA, 
which reinforces the B type, constrains the spectral type of LS~II~+26~8 to B0, which is also in 
agreement with the presence and relative intensity of Si~\textsc{iii} lines at 4552-68-75 ~\AA, 
the O~\textsc{ii} lines at 4070-76~\AA, Si~\textsc{iv} at 4089~\AA~and, in particular, 
the presence of very weak He~\textsc{ii} at 4200~\AA~and weakening He~\textsc{ii} at 4686~\AA\ 
 \citep{liu19}. The luminosity of the star was defined, also following \citet{liu19} 
(see their Table\,2), taking into account the presence of the O~\textsc{ii} line at 4380~\AA, 
the N~\textsc{ii} line at 3995~\AA~and, in particular, the more intense He~\textsc{i} line at 4713~\AA\  
compared to the He~\textsc{ii} line at 4686~\AA. Based on these arguments, the 
spectral and luminosity classification of LS~II~+26~8 is derived to be B0 III.

Therefore, we conclude that the star analysed is indeed an evolved massive star. Moreover, a 
careful inspection of the 24~$\mu$m emission in the nearby neighbourhood of the star supports this scenario. 

\begin{table}
\centering
\caption{Relevant absorption lines identified in Fig.\,\ref{fig:espectro} with their equivalent widths.}
\label{tab:eqwidth}
\begin{tabular}{l c c c l c c }
\hline
Sp. line & $\lambda\lambda$ &EW & &Sp. line & $\lambda\lambda$ &EW\\
 & \AA~ &\AA~  & & & \AA~ &\AA~\\
\hline
He~\textsc{i} & 3820 & 0.8336& &Si ~\textsc{iii} & 4552  &0.3213\\
               & 3927 & 0.5039 &&& 4568 & 0.2588 \\
               & 4009 & 0.3752&  && 4575 & 0.1723\\
		& 4026     & 0.8601 & & Si ~\textsc{iv} & 4089  &0.2816\\
		               & 4121 & 0.6157& && 4116 & 0.1955\\
		& 4144 & 0.4746 & &N ~\textsc{ii} & 3995 & 0.1841\\
                & 4169  & 0.0571 &  &N ~\textsc{iii}& 4379 & 0.1151\\
		& 4388 & 0.5187&  && 4640 & 0.1544\\ 
		& 4471 & 0.8226 & & & 4642 & 0.4998\\
		&4713 & 0.2951& &O ~\textsc{ii} & 4070 & 0.2537\\
		& 4924 & 0.6457 & && 4076 & 0.2834\\
He~\textsc{ii} & 4200 & 0.0351 &   &&  4350 & 0.3118\\
& 4542 & 0.0245  & && 4416 & 0.3807\\
&4686 & 0.1644 & 	&& 4676 & 0.1968		\\  	
Mg~\textsc{ii} & 4481 & 0.1849&&&&\\
\hline
\hline
\end{tabular}
\end{table}

\begin{figure}
    \centering
	\includegraphics[width=8.3cm]{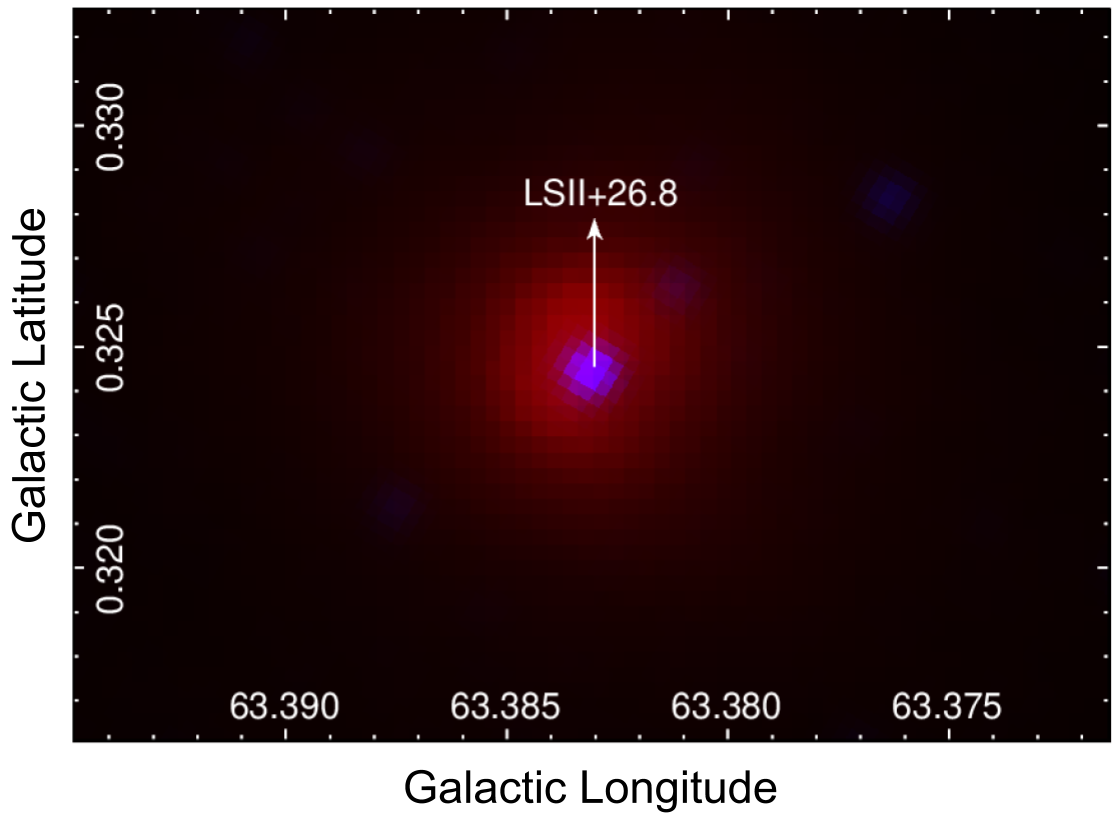}
    \caption{MIPS/{\it Spitzer} emission at 24~$\mu m$ (red) and the {\it J} 
broad-band emission (blue) from 2MASS towards the star LS~II~+26~8.}
    \label{fig:LSII26}
\end{figure}

Figure\,\ref{fig:LSII26} shows a two-colour image of the 24~$\mu$m emission extracted 
from the MIPS/{\it Spitzer} survey presented in red, and the {\it J} broad-band emission 
obtained from 2MASS in blue towards LS~II~+26~8. 
The star observed at the  near-IR {\it J} band is located at the centre of a bubble as 
seen at 24~$\mu$m. This source can therefore be identified as a MIPSGAL bubble (MB) (e.g.\  
\citealt{mizuno10,flag14}). As those authors indicate, most MBs are thought to be 
associated with stars, some of them massive, in the late stages of evolution. 
In those cases, the just-mentioned bubble morphology can be produced as a consequence of the star´s 
evolution: as its atmosphere expands and cools, the ejected gas condenses into dust grains 
within a circumstellar shell that emits at mid-IR wavelengths. The morphology of the 
24~$\mu$m emission observed in LS~II~+26~8 is quite similar to that observed towards 
MB 3701 generated by MN95, a \textsc{B0 I} star \citep{flag14}, and towards MB 3955 
\citep{nowak14}, which is generated by the star CD-61 3738, a likely OBe-type star \citep{steph71}.

According to \citet{hohle10}, the bolometric luminosity of a B0 III type star can be (1.60--2.06) $\times 10^{4}$~L$_{\odot}$. 
Using this value and the magnitude $m_{V} = 12.14$ \citep{zaca12} in the typical equation:

\begin{equation}
\log(L_{\rm bol}/L_\odot)=0.4(5\log(d/10\,{\rm pc})+4.8 - BC - m_{V} + A_{V}) 
\label{eq.lum}
\end{equation}

\noindent with the bolometric correction $BC = -2.8$ for a B0 star and a visual 
extinction  $A_{V} = 2.91$ (from $R_{V}=A_{V}/E(B-V)$, with $E(B-V)=0.97$,  
estimated using the $V$ and $B$ magnitudes of \citealt{zaca12}, the intrisic colour 
from \citealt{Wegner94}, and $R_{V}=3.0$), we derive a distance to the star in a range 
between 2.68 and 3.04 kpc.  
This calculation of the distance is totally independent of the distance estimates discussed in Sect.\,\ref{sec.present}.
Thus, we can confirm that LS~II~+26~8 and the \hii~region complex are indeed located at the same distance.

\subsection{The ISM around LS~II~+26~8 on a large spatial scale}

With the aim of finding evidence supporting the hypothesis that the massive evolved star
 LS~II~+26~8 has swept up molecular gas that probably gave rise to the stars exciting the
 \hii~region complex shown in Fig.\,\ref{fig:Sh2-90}, we investigate the ISM 
surrounding the star on a larger spatial scale. Unfortunately, at these Galactic 
latitudes/longitudes there is no molecular line survey covering the whole region. 
However, from the mid- and far-IR bands obtained from the {\it Herschel}-Hi-GAL survey
\citep{molinari10} it is possible to derive an H$_{2}$ column density map that 
indicates the distribution of molecular gas around LS~II~+26~8. 

\begin{figure}
    \centering
	\includegraphics[width=8.3cm]{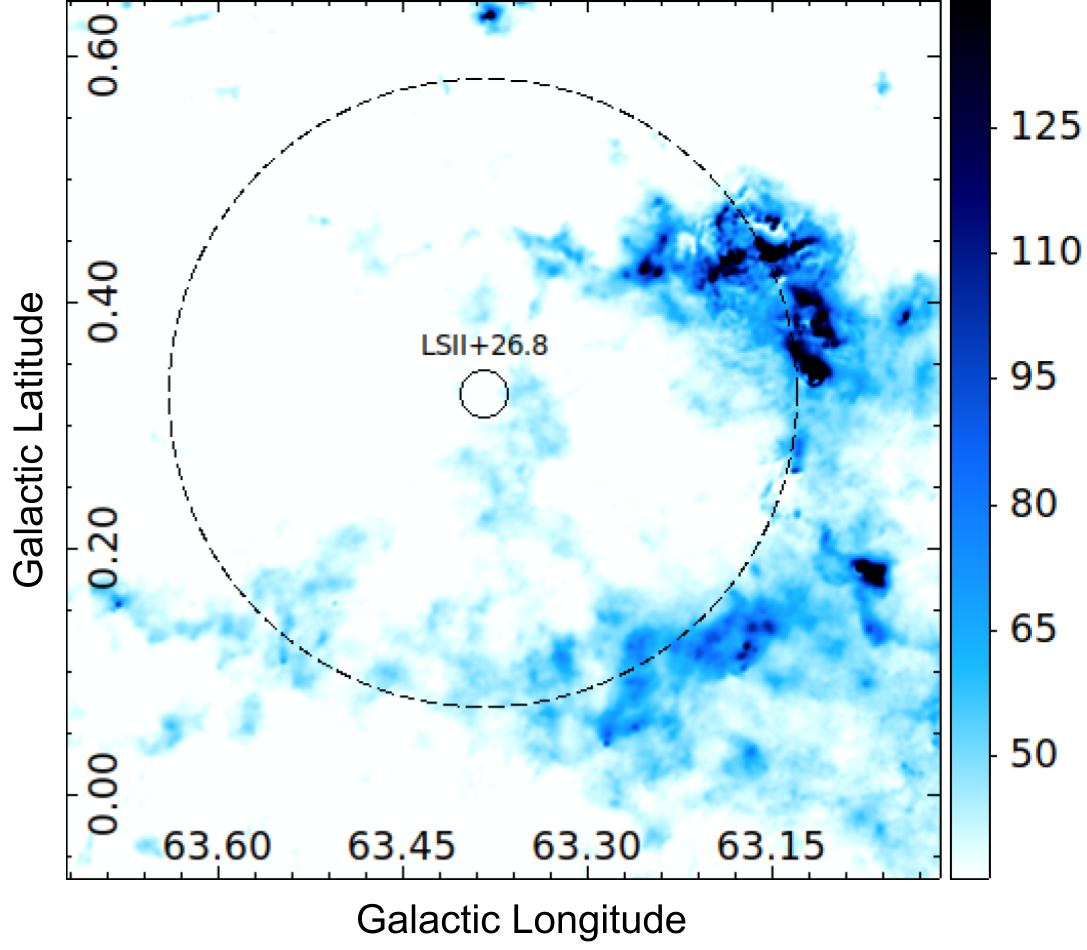}
    \caption{Distribution of the H$_2$ column density, N(H$_2$), derived from the 
PPMAP procedure applied to the {\it Herschel}-Hi-GAL data. The colour bar is in units
 of $\times10^{20}$ cm$^{-2}$.}
    \label{fig:NH2}
\end{figure}

We obtained an H$_{2}$ column density map generated from the PPMAP procedure carried out on the 
Hi-GAL maps in the wavelength range 70--500 $\mu$m (see 
\citealt{marsh17}).\footnote{http://www.astro.cardiff.ac.uk/research/ViaLactea/} 
Figure\,\ref{fig:NH2} shows the distribution of the H$_2$ column density. It can be
 seen that the star LS~II~+26~8 is located exactly at the geometric centre of a large 
H$_{2}$ open shell in whose northwestern/western border lies the above-described 
\hii~region complex. By inspecting the catalogues of massive/luminous stars in the 
VizieR and SIMBAD Astronomical Database we find that LS~II~+26~8 is the only massive 
star that lies within the molecular shell.
Some weaker N(H$_{2}$) features are observed within the circumference drawn by the 
shell, which could be due to material on the back and/or front borders of the 
probably expanding shell. This scenario supports the hypothesis that an extinct H\textsc{ii}~region excited 
by LS~II~+26~8 and/or its stellar winds could have generated the observed shell in 
which the massive stars responsible of Sh2-90 and G63 were born. 

\subsection{Association between LS~II~+26~8 and the molecular shell}

It is known that a massive star loses a large amount of mass in the form of stellar wind. 
The winds expand into the surrounding medium and collide with the gas in the ISM, 
generating a low density bubble that expands over time \citep{van15,mckee84}.

The location of LS~II~+26~8 exactly at the geometric centre of a molecular 
semi-shell, strongly suggests a causal connection between both, so we wondered whether 
this star was able to create such a molecular structure in the ISM.
Massive stars interact with the interstellar medium throughout their entire evolution. 
Hence, we have to take into account the possibility that, before reaching the B0 III stage, LS~II~+26~8 had 
evolved from a hotter main sequence star of an earlier spectral type. 
Thus, using the evolutionary tracks for massive stars of \citet{eks12} and the luminosity 
value of about $1.83\times10^{4}$ L$_{\odot}$ for a B0 III type star mentioned above, 
LS~II~+26~8 lies somewhere between the 12--15 M$_{\odot}$ evolutionary tracks, which correspond to an O9V.

Using the equation presented in \citet{cheva99} for a wind-blown bubble radius ($R_{B}$):
\begin{multline}
R_{B} = 15.8\left(\frac{\dot{M}}{10^{-8}~\rm{M_{\odot}~yr^{-1}}}\right)^{1/3} \left(\frac{v_{\rm w}}{10^{3}~\rm{km~ s^{-1}}}\right)^{2/3} \\
\times \left(\frac{\tau_{\rm ms}}{10^{7}~\rm{yr}}\right)^{1/3} \left(\frac{p/k}{10^{5}~\rm{K~cm^{-3}}}\right)^{-1/3}    
\label{bubble}
\end{multline}

\noindent with a time in the main sequence of $\tau_{ms} \sim 10^{7}$ yr, a mass loss 
rate $\dot{M} \sim 6 \times 10^{-8}$ M$_{\odot}$~yr$^{-1}$, 
and a terminal wind velocity $v_{\rm w} \sim 1300$ km~s$^{-1}$~(values corresponding to an 
O9V star, see \citealt{kobul19,kobul18,chen13,cheva99}), and assuming the typical 
interstellar medium pressure of $p/k = 10^{5}~\rm{K~cm^{-3}}$, we obtain $R_{B}$ about 13 pc. 

Assuming a distance about 2.8 kpc for the region, the radius of the molecular semi-shell is 
$\sim$12.2 pc, which is in agreement with the radius of the wind-blown bubble estimated 
above, thus supporting the hypothesis that LS~II~+26~8 is responsible of the molecular semi-shell presented in Fig.\,\ref{fig:NH2}.
Additionally, considering that LS~II~+26~8 is a \textsc{B0 III} star, its age should be
 $\gtrsim 10^{7}$ yr, whereas Sh2-90 has an age $< 6.5 \times 10^{6}$ yr \citep{2014A&A...566A.122S}. 
A comparison of these times gives support to the proposed scenario of probable triggered star formation.

\section{Summary and conclusion}

Evidence for triggered star formation linking several generations of stars is difficult to assemble. 
Based on a detailed study of the star LS~II~+26~8 and the surrounding ISM, we present a plausible
case for triggered star formation that links three generations of stars. We focus on LS~II~+26~8 
because it was catalogued as a possible OB-type star, and a semi-shell-like \hii~region complex,
 which in turn presents signs of triggered star formation, that seems to be related to it.  
Our main results are as follows:

(a) From our spectroscopic optical observations carried out with the Isaac Newton Telescope 
at Roque de Los Muchachos Observatory, we classified LS~II~+26~8 as a B0 III, i.e.\ an evolved massive star.

(b) Analysing the mid-IR emission towards LS~II~+26~8, we note that the star is embedded 
in a bubble that emits at 24 $\mu$m, thus supporting the case that LS~II~+26~8 is a massive star that has left the main sequence. 

(c) Using a range of bolometric luminosity corresponding to a B0 III type star and the optical
 magnitudes measured for this star, we determine that LS~II~+26~8 is located at a distance 
of about 2.8 kpc. Additionally, by following evolutionary tracks for massive stars we note that 
during its life on the main sequence the spectral type of the star should have been O9.

(d) We observe that LS~II~+26~8 is located exactly at the geometric centre of a semi-shell-like 
\hii~region complex (15\m~in radius). It is known that the  \hii~region Sh2-90, which belongs 
to this complex, has more than 100 YSO candidates in its vicinity. The estimated distances 
to LS~II~+26~8 and to Sh2-90 are in agreement.

(e) Analysing the distribution of the H$_2$ column density obtained from Hi-GAL maps in the 
wavelength range 70--500 $\mu$m, we observe an extended shell also of 15\m~in radius on whose 
northwestern border lies the \hii~region complex. LS~II~+26~8 is located exactly at its geometric 
centre and is the only catalogued massive star within the shell.

(f) Assuming a distance about 2.8 kpc to the region, the radius of the molecular shell is 
$\sim13$ pc, which is in agreement with what an O9V star can generate through its winds in 
the molecular environment. 

Based on the above points, we conclude that we have presented strong evidence pointing to a 
probable triggered star-forming scenario. The action of LS~II~+26~8 could have triggered the 
formation of the exciting stars of the \hii~region complex at the shell, which in turn, is
very probably triggering a new generation of stars.

\section*{Acknowledgements}

We thank the anonymous referee for her/his very helpful comments and suggestions. We are specially grateful to Terry Mahoney for the language revision of the manuscript.
M.B.A. is a doctoral fellow of CONICET, Argentina.
A.B., S.P. and  M.O. are members of the {\sl Carrera del Investigador Cient\'ifico} of CONICET, Argentina.
The INT is operated on the island of La Palma by the Isaac Newton Group in the Spanish Observatorio del 
Roque de los Muchachos of the Instituto de Astrof\'isica de Canarias.



\bibliographystyle{mnras}
\bibliography{ref.bib} 








\bsp	
\label{lastpage}
\end{document}